\documentclass[sigconf]{acmart}
\AtBeginDocument{%
  }

\copyrightyear{2026}
\acmYear{2026}
\setcopyright{cc}
\setcctype{by}
\acmConference[RecSys '26]{20th ACM Conference on Recommender Systems}{September 27-October 02, 2026}{Minneapolis, MN, USA}
\acmBooktitle{20th ACM Conference on Recommender Systems (RecSys '26), September 27-October 02, 2026, Minneapolis, MN, USA}
\acmDOI{10.1145/3773078.3831945}
\acmISBN{979-8-4007-2284-4/2026/09}

\usepackage{graphicx}
\usepackage{caption}
\usepackage{subcaption}
\usepackage{bm}
\usepackage{tabularx}
\usepackage{multirow}
\usepackage{makecell}
\usepackage{xr}
\usepackage{placeins}
\usepackage{enumitem}
\usepackage{array}
\usepackage{float}
\usepackage{dblfloatfix}
\usepackage{makecell}
\usepackage{amsmath}
\usepackage{algorithm}
\usepackage{algpseudocode}
\usepackage{tikz}
\usepackage{xcolor}
\usepackage{pdfpages}
\usepackage{hyperref}       
\usepackage{url}            
\usetikzlibrary{arrows.meta, positioning}

\newcolumntype{C}[1]{>{\centering\arraybackslash}m{#1}}
\fontsize{7.8}{8}\selectfont

\begin{document}


\newcommand{\keywordList}{recommender systems, learning to rank, collaborative filtering, graph neural networks, social recommendation, online content platforms, position bias, popularity bias, unbiased recommendation}
\title{Unbiased Recommender Systems with Implicit Feedback}


\author{Md Aminul Islam}
\affiliation{
  \institution{University of Illinois Chicago}
  \city{Chicago}
  \state{IL}
  \country{USA}
}
\email{mislam34@uic.edu}

\renewcommand{\shortauthors}{Md Aminul Islam}

\begin{abstract}

Recommender systems typically rely on implicit feedback (e.g., clicks) to infer user preferences. However, such data is inherently prone to various biases, including position bias and popularity bias. Position bias occurs when higher-ranked items receive more interactions regardless of true relevance. Popularity bias reinforces frequent exposure of popular items while under-recommending relevant, yet less popular ones. Directly learning from such data fails to capture true user preferences, leading to suboptimal recommendations. This research focuses on mitigating position bias and popularity bias in recommender systems. Specifically, I address position bias in learning-to-rank (LTR) systems and popularity bias in collaborative filtering (CF) models and social recommender systems based on graph neural networks. My work develops methods that overcome the limitations of existing approaches to mitigating position bias and popularity bias, enabling more relevant and personalized recommendations that align with users’ preferences.    
\end{abstract}


\ccsdesc[500]{Information systems~Learning to rank}
\ccsdesc[500]{Information systems~Personalization}
\ccsdesc[500]{Information systems~Social recommendation}

\keywords \keywordList


\maketitle

\section{Introduction} \label{sec:introduction}

Recommender systems have become a fundamental component of modern online platforms, enabling users to efficiently navigate large-scale information spaces and discover personalized, relevant content. They are widely used across a wide range of complex digital ecosystems, including e-commerce, entertainment services, search engines, social networks, and lifestyle applications~\cite{chen-inf23}. Recommender systems learn user–item relevance from historical interaction data and produce ranked lists that prioritize items most aligned with each user’s inferred preferences~\cite{wu-springer10}. They commonly rely on implicit feedback data, such as user clicks, to infer user preferences, as this type of data is easy to collect from interaction logs~\cite{saito-wsdm20} and serves as a valuable signal of user behavior~\cite{sanderson-ir10}. However, in real-world scenarios, implicit feedback is not only driven by true user preferences but is also influenced by various exposure mechanisms~\cite{luo-wsdm23}. For example, a click can depend on what the system chooses to expose~\cite{oosterhuis-sigir20, ovaisi-www20}, where the item is placed~\cite{ai-sigir18, joachims-wsdm17}, how frequently it appears~\cite{chen-front24, zhou-sigir23}, and how prior recommendations shape future behavior~\cite{klimashevskaia-umuai24}. These factors collectively introduce systematic biases into the observed data~\cite{ai-sigir18, joachims-wsdm17, yue-www10}. As a result, models trained directly on such biased implicit feedback learn distorted preference signals, leading to suboptimal recommendations and reduced personalization~\cite{joachims-wsdm17}. These biases can also result in feedback loops, where items that receive more exposure or interactions are increasingly promoted in future recommendations~\cite{chaney-recsys18}. Over time, this feedback loop progressively reinforces and amplifies existing biases, resulting in a “rich-get-richer” effect~\cite{klimashevskaia-umuai24}, which ultimately degrades the long-term performance of recommender systems.

\begin{figure} [b]
  \centering
  \captionsetup{justification=raggedright, margin=0cm}
  \vspace{-5pt}

  \includegraphics[width=0.46\textwidth, alt="A two panel figure showing that higher ranked items receive higher interaction probability due to position bias while a small set of popular items dominates interactions in a long tail distribution due to popularity bias."]{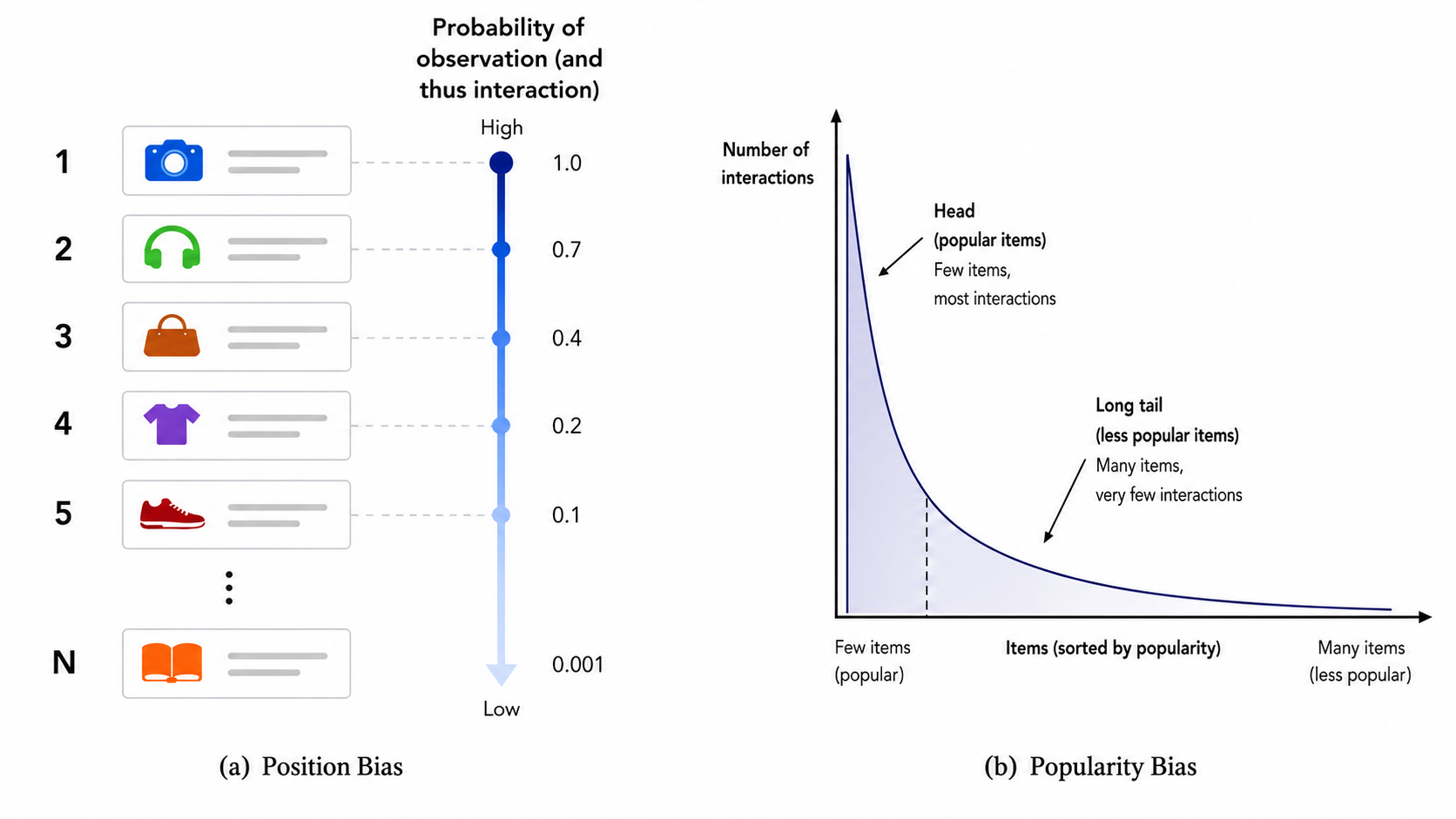}

  \vspace{-4pt}
  \captionsetup{width=0.48\textwidth}
  \caption{(a) Position bias: Higher-ranked items are more likely to be observed and clicked, regardless of true relevance. (b) Popularity bias: Popular items accumulate most interactions, causing models to overestimate them and under-represent less popular but relevant items.}
  \label{fig:pos_pop_bias}
\end{figure}

Two of the most prevalent biases in recommender systems are position bias and popularity bias. Position bias arises when users are more likely to interact with items ranked at higher positions, regardless of their true relevance~\cite{chen-inf23, craswell-wsdm08, joachims-tois07}. Models trained directly on such biased data tend to overestimate the relevance of higher-ranked items while underestimating lower-ranked but potentially relevant ones~\cite{ai-sigir18, joachims-wsdm17}. Popularity bias arises when the distribution of observed implicit feedback is skewed toward a small set of popular items~\cite{zhang-neurips22}, causing models trained on this data to inherit popularity as a proxy for true relevance~\cite{canamares-sigir18, yao-neurips17}. Consequently, relevance is overestimated for popular items and underestimated for niche but relevant items~\cite{boratto-ipm21}. Figure~\ref{fig:pos_pop_bias} illustrates these two types of bias.

Learning-to-rank (LTR) and collaborative filtering (CF) are two widely used paradigms in recommender systems.  An LTR system aims to learn a ranking model that ranks items according to their relevance to a 
user query~\cite{liu-ftir09}, and is therefore especially affected by position bias in ranked feedback. CF models infer user preferences from historical user--item interactions by leveraging patterns of similar user behavior~\cite{lubos-frontiers23}. This makes CF effective for personalized recommendation, but also vulnerable to popularity bias when interactions are concentrated on popular items.

Existing position bias correction methods for LTR mainly include click models, propensity-based methods, and econometric approaches. Click modeling methods~\cite{guo-recsys19, yan-sigir22} attempt to separate bias from relevance, but can fail when position and relevance are strongly confounded~\cite{zhang-kdd23}. Propensity-based methods~\cite{joachims-wsdm17, ai-sigir18, luo-sigir24} use inverse propensity weighting (IPW) to obtain relevance-equivalent objectives, but they are sensitive to propensity misspecification~\cite{ovaisi-sigir21}, high variance~\cite{luo-wsdm23}, and often require randomization~\cite{joachims-wsdm17}. Heckman-based econometric methods~\cite{ovaisi-www20, ovaisi-sigir21, heckman-econometrica79} avoid propensity estimation, but their linearity assumptions limit their ability to model complex nonlinear feature interactions. Click modeling and propensity-based methods require modifying the ranker’s objective, making them tightly coupled to specific LTR algorithms. These limitations motivate a model-agnostic position bias correction method that does not require propensity estimation, better controls bias-related signals, and remains compatible with nonlinear ranking models. 

Prior work mitigates popularity bias for CF using IPW, re-ranking, regularization, and causal inference methods. IPW-based methods assign lower weights to interactions with frequent items~\cite{gruson-wsdm19}. Post-processing re-ranking methods~\cite{abdollahpouri-arXiv19, zhu-kdd21, zhu-wsdm21} mitigate popularity bias by adjusting the final recommendation lists. Regularization-based methods~\cite{boratto-ipm21, chen-sigir20} penalize the dependence between predicted scores and item popularity. Causal graph-based methods~\cite{ning-www24, wei-sigkdd21} model popularity bias through causal structures to identify and control its sources. While recommender systems based on graph neural networks (GNNs) achieve state-of-the-art performance in recommendation tasks~\cite{he-sigir20, gao-wsdm22}, they amplify popularity bias~\cite{zhou-sigir23, chen-front24}. In GNN-based recommender systems, popular items are connected to many users, allowing their signals to propagate to a large number of multi-hop neighbors during message passing. 
While existing approaches can reduce the dominance of popular items, they do not address how popularity bias becomes embedded in GNN node representations or how it is amplified through the GNN message-passing process. These limitations motivate popularity bias correction methods in GNN-based CF that address popularity amplification by removing popularity components from node representations or reducing popularity influence during GNN message passing.

The rapid growth of online social platforms has increased interest in social recommender systems, which integrate user--user connections with user--item interactions~\cite{sharma-acmcs24}. However, in user-generated content-sharing platforms (e.g., Facebook and TikTok), items are produced by creators who are also part of the social ecosystem. Existing content-sharing platform methods~\cite{chen-www25, ionescu-recsys23, li-www26} overlook social relations among users, while social recommender systems~\cite{fan-www19, wu-tkde22, wu-sigir19, xu-ijoc25, yu-www21} ignore creator-item information. This leaves an important gap in social recommender systems: content from frequently posting or highly popular creators can dominate user feeds. Moreover, existing social recommender systems do not account for how popularity bias can be amplified through GNN message passing. In social recommender systems for user-generated content platforms, popularity bias can arise from multiple sources: popular items receive more interactions, popular creators gain disproportionate exposure for their content, and highly influential users can exert disproportionate influence on their neighbors’ representations. This motivates the need for GNN-based social recommender systems for user-generated content platforms that jointly account for item, creator, and social-neighbor popularity.

All these challenges motivate the central research question of my PhD: \textit{How can reliable user preferences be learned when implicit feedback  is systematically biased (e.g., position bias, popularity bias)?} My work develops methods to reduce these biases in recommender systems by addressing the limitations of existing methods.

\textbf{Research contributions.}
To address these challenges, my PhD research makes the following key contributions:
\begin{itemize}[leftmargin=9pt, nosep]
\item \textit{Position bias correction in LTR systems:}
I develop a model-agnostic framework for correcting position bias in LTR systems~\cite{islam-recsys26} based on an econometric two-stage control function approach. The method does not require propensity estimation, supports nonlinear models, and can be integrated into existing LTR algorithms without modifying their ranking objectives. \textit{This work has been accepted for publication in the Proceedings of the ACM Conference on Recommender Systems 2026.}

\item \textit{Post-hoc popularity debiasing in GNN-based CF:}
I propose a post-hoc popularity debiasing method for GNN-based CF~\cite{islam-www26} that removes the popularity component from pre-trained node embeddings using interaction-level popularity, without requiring retraining. \textit{This work has been accepted for publication in the Proceedings of the ACM Web Conference 2026.}

\item \textit{Popularity-aware message passing in GNN-based CF:}
I develop a GNN-based debiasing method that mitigates popularity amplification during GNN message passing. It adaptively re-weights messages with interaction-level weighting and layer-wise weighting during message passing to reduce the dominance of popular items. \textit{This work is currently under review.}

\item \textit{Popularity bias correction in social recommender systems:}
As future work of the PhD, I plan to extend popularity-bias mitigation beyond traditional user--item CF models to GNN-based social recommender systems. This work aims to study how item popularity, creator popularity, and social-neighbor popularity jointly influence recommendations, and to reduce this multi-source popularity amplification during GNN message passing.

\end{itemize}

\section{Research Progress} \label{sec:research_progress}

My research has so far focused on directions that address position bias and popularity bias with implicit feedback. The first work addresses position bias in LTR systems, while the second and third works focus on popularity bias in GNN-based CF. The first work~\cite{islam-recsys26} and the second work~\cite{islam-www26} have been accepted to the ACM Conference on Recommender Systems 2026 and the ACM Web Conference 2026, respectively, while the third work is currently under review.

\subsection{Correcting for position bias in LTR systems using control function approach}

My first work addresses position bias in LTR systems. In LTR, users are more likely to observe and click higher-ranked items, even when lower-ranked items may be equally or more relevant. I propose Control Function-based Correction (CFC), a two-stage framework for correcting position bias in LTR systems. The control function method~\cite{wooldridge-jhr15} is a prominent approach for addressing different types of data bias in econometrics literature. In the first stage, this approach models the biased or endogenous variable and estimates residuals from the model. In the second stage, these residuals are included as control terms in the outcome model to account for bias.

In the first stage of CFC, I model the previous ranking process that generated the training data, where users were more likely to click on higher-ranked items. Item relevance depends on query--item features, and these features are used by the ranking policy to determine item positions. I model the historical ranking process using features as inputs and item rank as the target output. I then estimate the residual for each query--item pair as the difference between the observed rank and the predicted rank. These residuals represent unobserved factors that influence ranking outcomes beyond the observable features, such as system biases, limited or noisy training data, randomization, or personalization effects. They capture variation in item placement for a given query that is not explained by observable feature-driven relevance. In the second stage, I incorporate the estimated residuals to control for variation in the ranking process beyond what is explained by observable features. The residuals and their interactions with each feature are added as control function terms in the second stage click model.
Conditioning on these terms helps account for position-induced dependence in the second-stage model to correct for position bias.

CFC does not require propensity estimation or result randomization. It also does not require modifying the ranker’s objective function, making it compatible with different LTR algorithms. Moreover, the control function method does not impose the restrictive linearity assumptions of Heckman-style econometric approaches. Another practical contribution is a debiased validation-click strategy for hyperparameter tuning when true relevance labels are unavailable. Since validation clicks are affected by position bias, directly tuning on raw clicks can select a suboptimal ranker. To address this, I remove the position-induced component from validation clicks. 

I evaluate CFC on three LTR benchmark datasets and a large-scale industrial dataset. CFC improves ranking performance over existing position bias correction methods and across multiple metrics and nonlinear rankers. CFC remains robust to varying bias severity, noisy clicks, and different logging policies.

Overall, this work makes the following key contributions:
\begin{itemize}[leftmargin=9pt, nosep]
    \item I propose a control function-based method for correcting position bias in LTR systems.
    \item CFC is model-agnostic, supports nonlinear ranking models, and does not require propensity estimation or result randomization.
    \item I propose a residual-based click debiasing method for hyperparameter tuning when unbiased validation labels are unavailable.
    \item CFC outperforms state-of-the-art baselines on benchmark and industrial datasets, and is robust to different position-bias severities, noisy clicks, and deterministic and stochastic logging policies.
\end{itemize}

\subsection{Post-hoc popularity bias correction in GNN-based CF}

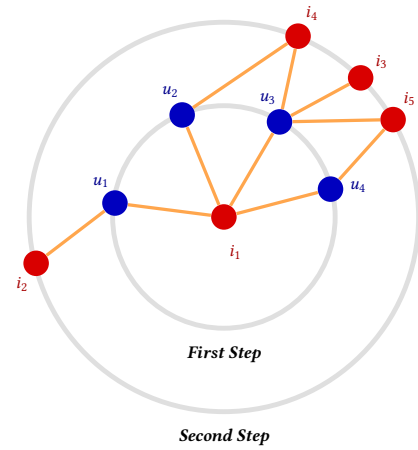
\begin{figure}[ht]
\centering
\captionsetup{justification=raggedright, margin=0cm}
\resizebox{0.31\textwidth}{!}{
\begin{tikzpicture}[
    user/.style={circle, fill=blue!75!black, minimum size=0.55cm, inner sep=0pt},
    item/.style={circle, fill=red!85!black, minimum size=0.55cm, inner sep=0pt},
    edge/.style={line width=2.0pt, draw=orange!70},
    labeluser/.style={font=\Large\bfseries\itshape, text=blue!55!black},
    labelitem/.style={font=\Large\bfseries\itshape, text=red!65!black},
    steplabel/.style={font=\Large\bfseries\itshape}
]

\draw[gray!25, line width=3pt] (0,0) circle (2.4cm);
\draw[gray!25, line width=3pt] (0,0) circle (4.2cm);

\node[steplabel] at (0,-2.95) {First Step};
\node[steplabel] at (0,-4.75) {Second Step};

\node[user] (u1) at (-2.35,0.3) {};
\node[user] (u2) at (-0.9,2.20) {};
\node[user] (u3) at (1.2,2.05) {};
\node[user] (u4) at (2.3,0.6) {};

\node[item] (i1) at (0,0) {};
\node[item] (i2) at (-4.05,-1.0) {};
\node[item] (i3) at (2.95,3.00) {};
\node[item] (i4) at (1.6,3.9) {};
\node[item] (i5) at (3.65,2.1) {};

\draw[edge] (u1) -- (i1);
\draw[edge] (u2) -- (i1);
\draw[edge] (u3) -- (i1);
\draw[edge] (u4) -- (i1);

\draw[edge] (u1) -- (i2);
\draw[edge] (u3) -- (i3);
\draw[edge] (u2) -- (i4);
\draw[edge] (u3) -- (i4);
\draw[edge] (u3) -- (i5);
\draw[edge] (u4) -- (i5);

\node[labeluser] at (-2.65,0.75) {$u_1$};
\node[labeluser] at (-1.15,2.70) {$u_2$};
\node[labeluser] at (0.95,2.55) {$u_3$};
\node[labeluser] at (2.90,0.65) {$u_4$};

\node[labelitem] at (0.25,-0.75) {$i_1$};
\node[labelitem] at (-4.35,-1.45) {$i_2$};
\node[labelitem] at (3.40,3.35) {$i_3$};
\node[labelitem] at (1.90,4.40) {$i_4$};
\node[labelitem] at (4.00,2.55) {$i_5$};

\end{tikzpicture}
}
\vspace{-4pt}
\caption{An illustrative example of popularity bias in GNN-based CF. The popular item $\bm{i_1}$ is connected to many users, so within two propagation steps it aggregates information from a much larger neighborhood and spreads its signal to more nodes. In contrast, the niche item $\bm{i_2}$ has fewer connections and receives limited neighborhood information. As a result, popular-item signals become overrepresented in the learned embeddings.}
\label{fig:two_step_graph}
\end{figure}

My second work addresses popularity bias in GNN-based CF. GNNs amplify popularity bias through their message passing process~\cite{zhou-sigir23, chen-front24}. In GNN-based CF, users and items are represented as nodes in a user--item interaction graph, and their embeddings are learned by repeatedly aggregating information from neighboring nodes. During each message-passing layer, each node aggregates information from its neighbors’ representations in the previous layer and captures high-order neighborhood information through repeated propagation. Since popular items are connected to many users, their signals can spread more widely to higher-order neighborhoods during repeated message passing. This can move many user embeddings closer to popular items in the embedding space, increasing the chance that popular items receive higher prediction scores. Therefore, the popularity effect can be amplified through GNN message passing and neighborhood aggregation. Figure~\ref{fig:two_step_graph} shows how popular items aggregate and spread signals to more nodes than niche items during GNN message passing.

Existing methods often mitigate popularity bias by modifying training objectives, but they do not directly remove bias embedded in node representations during GNN message passing. Recent GNN-based aggregation-weighting methods downweight popular-item interactions during message passing using either static weights from interaction counts~\cite{kim-cikm22} or dynamic weights estimated from embeddings of the evolving model~\cite{zhou-sigir23}. However, static weights may fail to capture model-specific representation-aware popularity signals, while dynamic weights can be unstable because the embeddings are still evolving during training. A recent post-hoc method~\cite{chen-front24} for GNN-based CF estimates popularity effects using simple node degrees, but it does not consider personalized preferences or interaction-level popularity effects.

To address these limitations, I propose Post-hoc Popularity Debiasing (PPD), a post-hoc method for mitigating popularity bias in GNN-based CF. PPD operates directly on pre-trained user and item embeddings and does not require retraining the model. The key idea is to estimate how much each user--item interaction is influenced by popularity, construct a popularity direction vector in the embedding space, and remove the popularity component from node embeddings through vector projection.

PPD first estimates the popularity effect of each user--item interaction using the embeddings of a pre-trained model. For this, it considers both the global preference of an item and the personalized preference of a user for that item. Global preference captures how broadly an item is preferred across users, while personalized preference captures how well the item aligns with the user's historical interests. The interaction-level popularity score is then estimated by removing the personalized preference from global preference. The popularity score reflects how much the interaction is influenced by popularity alone. After that, PPD constructs a popularity direction vector for each node. For each user or item node, the method computes two neighbor-based embeddings: a popularity-based embedding and a preference-based embedding. The popularity-based embedding is obtained by weighting interacted neighbors with their popularity scores, while the preference-based embedding is obtained by weighting them with the complement of the popularity scores. The difference between these two defines the popularity direction of a node, which represents how popularity pulls the node embedding away from preference-driven information. PPD then projects each pre-trained node embedding onto this popularity direction and subtracts the projected component. This removes the popularity-aligned part of the pre-trained embedding while preserving preference-related information.

I evaluate PPD on three real-world datasets with unbiased test data. The results show that PPD outperforms state-of-the-art popularity bias correction methods from different methodological families across multiple recommendation metrics. PPD also improves recommendation quality for both popular and niche items, showing that reducing popularity bias does not require sacrificing performance on genuinely relevant popular items. The method performs consistently across different GNN layer depths, generalizes to another GNN-based CF backbone, and is computationally inexpensive.

Overall, this work makes the following key contributions:
\begin{itemize}[leftmargin=9pt, nosep]
    \item I propose a post-hoc popularity bias correction method for GNN-based CF, which does not modify the training objective, making it applicable to any GNN-based CF model.
    \item PPD operates on pre-trained embeddings and does not require retraining, allowing it to be applied to already deployed models.
    \item PPD operates directly on learned embeddings, allowing popularity estimation from stable representations.
    \item PPD estimates popularity at the interaction level to capture finer-grained popularity effects, incorporating both global and personalized influences in the popularity estimation.
    \item PPD outperforms state-of-the-art popularity-bias correction baselines, improves recommendation performance for both popular and niche items, remains effective across different GNN layer depths, and is computationally efficient.
\end{itemize}
\vspace{-8pt}

\subsection{Debiasing message passing to mitigate popularity
bias in GNN-based CF}

My third work also addresses popularity bias in GNN-based CF, but unlike the previous post-hoc method, this work intervenes directly during GNN message passing. The motivation is that popularity bias can also be amplified during neighborhood aggregation.

Recent aggregation-weighting methods for GNN-based CF reduce the influence of popular-item interactions during message passing by using either interaction-count-based static weights~\cite{kim-cikm22} or embedding-based dynamic weights from the evolving model~\cite{zhou-sigir23}. However, static weights may overlook popularity signals encoded in learned representations, while dynamic weights from evolving embeddings can be noisy and unreliable during training. 

To address these limitations, I propose Debiasing Popularity Amplification in Aggregation (DPAA), a popularity-debiasing framework for GNN-based CF that directly modifies the message-passing process. DPAA assigns popularity-aware weights to user--item interactions during aggregation, so that interactions likely to reflect popularity-driven signals receive smaller weights, while interactions involving less popular but potentially relevant items receive relatively larger weights. In this way, DPAA reduces the dominance of popular items during message passing.

The first component of DPAA is an interaction-level weighting mechanism. The method estimates the strength of each user--item interaction using the similarity between user and item embeddings. Since popular items tend to become broadly aligned with many users, strong user--item similarity may reflect popularity-driven reinforcement rather than true user preference.  DPAA therefore uses the complement of this similarity as an inverse interaction weight, assigning lower weights to interactions that are likely to be popularity-driven. However, embeddings from the current model can be unstable, especially in the early stages of training. To address this, DPAA combines two sources of information for weight estimation: stable embeddings from a pre-trained model and embeddings from the current evolving model. Early in training, the method relies more on the pre-trained embeddings because they provide stable estimates. As training progresses and the current model becomes more reliable, the weighting gradually shifts toward the current embeddings. This smooth transition stabilizes message weights while still allowing the model to adapt during training. The second component of DPAA is layer-wise weighting. Lower GNN layers mainly capture immediate user--item interactions, which are often dominated by popular items. Higher layers aggregate information from multi-hop neighborhoods and can capture broader collaborative signals, including signals from diverse and underexposed items. DPAA assigns larger weights to deeper layers to increase the influence of higher-order neighborhood information. This moves the representations beyond immediate popularity-driven interactions.

I evaluate DPAA on three real-world datasets with unbiased test data and on semi-synthetic datasets with controlled popularity-bias severity. The results show that DPAA outperforms state-of-the-art popularity-bias correction methods from different methodological families across multiple metrics. DPAA also improves recommendation performance for both popular and niche item groups, showing that the method improves niche-item recommendation while preserving strong performance on popular items. The method remains effective across different levels of popularity-bias severity, with especially strong gains under high-bias settings.

Overall, this work makes the following key contributions:
\begin{itemize}[leftmargin=9pt, nosep]
    \item I propose a unified popularity-debiasing framework for GNN-based CF that directly mitigates popularity amplification during message passing by combining stabilized adaptive embedding-aware bias-correcting weights with layer-wise weights.
    \item DPAA stabilizes interaction-level weight estimation by smoothly combining embeddings from a pre-trained model and the current evolving model during training.
    \item I develop a semi-synthetic data generation process to systematically evaluate methods under different popularity-bias severities.
    \item DPAA outperforms existing popularity bias correction methods, improves both popular-item and niche-item performance, and remains effective across different levels of popularity bias.
\end{itemize}
\vspace{-6pt}
\section{Future Direction} \label{sec:future_direction}
Recommender systems for user-generated content platforms, such as Facebook~\cite{evnine-kdd24}, TikTok~\cite{chai-recsys25}, and YouTube~\cite{zhao-recsys19}, help users discover content produced by creators who are also part of the platform. These ecosystems typically involve three interconnected components: users who consume content, creators who produce content, and social relations that shape how preferences and recommendations propagate across the platform. However, popular content and content from popular creators often receive higher exposure, accumulate more interactions, and are repeatedly favored over time. This feedback loop further increases their visibility, while relevant content from new, niche, or less popular creators remains underexposed. As a result, the platform may overlook relevant but underexposed content, overshadow content from less popular yet socially relevant strong-tie peers, and reduce user satisfaction. 

Social recommender systems incorporate user--user relations, such as friendship, trust, or social interactions, into preference learning to alleviate data sparsity and improve performance~\cite{guo-aaai15, jamali-recsys10, ma-cikm08}. Graph-based methods further propagate signals over user--item and user--user graphs, allowing users to learn from social neighbors and higher-order interaction neighborhoods~\cite{fan-www19, wu-tkde22, wu-sigir19, yu-kdd21, yu-www21}. However, this propagation can also amplify popularity bias by repeatedly spreading signals from popular items and influential users~\cite{he-nn25, sheth-tkdd23, xu-ijoc25}. Thus, popularity bias in social recommender systems arises from user--item interactions and social influence.

Relatively few studies have addressed popularity bias in social recommender systems and content-platform recommender systems. Bias-aware social recommendation methods disentangle user interest, item popularity, and social influence~\cite{sheth-tkdd23}, adjust item-side social preferences~\cite{he-nn25}, or regulate social influence through  counterfactual approaches~\cite{wang-neucom25, xu-ijoc25}. Recommender systems for content platforms emphasize optimizing both user-side utility and creator-side outcomes such as exposure, diversity, fairness, and long-term ecosystem health~\cite{chen-www25, ionescu-recsys23, li-www26, patro-www20, xiao-kdd19}. However, most work focuses on exposure allocation~\cite{xiao-kdd19}, provider fairness~\cite{ionescu-recsys23, patro-www20}, or platform-level ecosystem design~\cite{chen-www25, li-www26}, with less attention to how creator popularity is amplified through user--item interactions, social relations, and graph propagation in GNN-based social recommendation.

Existing content-platform recommendation methods consider creator-side outcomes, but they do not directly address how popular creators may overshadow relevant content from less popular creators, nor do they model user--user social relations. Conversely, GNN-based social recommender systems leverage social connections among users, but they do not account for creator information, which can lead to the content of frequently posting, popular creators overwhelming user news feeds. Not accounting for such information can amplify popularity bias in social recommender systems. Moreover, existing bias-correction methods in social recommender systems do not account for how popularity bias can propagate through user--item and social relations during the GNN message passing. These gaps motivate a unified framework that jointly models user--item interactions, user--user social relations, and creator--item associations, while mitigating popularity amplification from items, creators, and social neighbors in GNN-based social recommender systems.

In this future work, I plan to study popularity bias in GNN-based social recommender systems, where users consume content, creators produce content, and social relations connect users or creators within the same platform. In this setting, popularity bias can arise from multiple sources: popular items accumulate more interactions, popular creators obtain more opportunities for their content to be exposed, and highly influential users can disproportionately influence their neighbors' representations through message passing. As a result, user--item and user--user graph propagation may repeatedly amplify signals from popular items, structurally dominant creators, and influential social neighbors. This can bias learned user representations and final rankings toward already popular content, rather than content that better reflects users' personalized preferences, including relevant content from niche creators or low-popularity strong-tie peers. To this end, I aim to propose a popularity-aware GNN framework for social recommender systems that preserves useful item, creator, and social signals while preventing popularity from dominating graph propagation. Unlike prior methods that separately focus on item-side bias, social influence bias, noisy social relations, or platform-level creator fairness, this work examines how popular items, popular creators, and highly connected users jointly shape exposure, representation learning, and recommendation outcomes through graph propagation.
\section{Conclusion} \label{sec:conclusion}
This research advances the development of unbiased recommender systems with implicit feedback by addressing position bias and popularity bias across multiple recommendation settings. First, it introduces a control function-based framework for correcting position bias in LTR systems. Second, it proposes a post-hoc method for mitigating popularity bias in GNN-based CF by removing popularity-driven components from learned embeddings. Third, it develops a message-passing debiasing framework that directly reduces popularity amplification during GNN message passing. Finally, I aim to extend this line of work to GNN-based social recommender systems, where item popularity, creator popularity, and social-neighbor popularity can jointly reinforce popularity bias and distort recommendation outcomes. Together, this research advances the development of unbiased recommender systems from implicit feedback across diverse settings, including LTR systems, personalized GNN-based CF, and social recommender systems.

\begin{acks}
I would like to thank my advisor, Elena Zheleva, along with my collaborators Kathryn Vasilaky, Ren Wang, Sourav Medya, and Ahmed Sayeed Faruk, for their guidance and support on my research.
\end{acks}

\bibliographystyle{ACM-Reference-Format}
\balance
\bibliography{main}
\balance


\end{document}